\documentclass[%
 aps,
 pre,
 amsmath,amssymb,
 reprint,%
]{revtex4-1}

\usepackage{graphicx}
\usepackage{dcolumn}
\usepackage{bm}

\usepackage[utf8]{inputenc}
\usepackage[T1]{fontenc}
\usepackage{etoolbox}
\usepackage{amsmath}
\usepackage{color}
\def\vector#1{\mbox{\bm{$#1$}}}
\newcommand{\XE}{\xi_{\rm e}}
\newcommand{\XI}{\xi_{\rm i}}
\newcommand{\XIA}{\xi_{\rm 1}}
\newcommand{\XIB}{\xi_{\rm 2}}
\newcommand{\XJ}{\xi_{\rm j}}
\newcommand{\NE}{n_{\rm e}}
\newcommand{\NI}{n_{\rm i}}
\newcommand{\NJ}{n_{\rm j}}
\newcommand{\te}{T_{\rm e}}
\newcommand{\ti}{T_{\rm i}}

\newcommand{\tia}{T_{\rm i1}}
\newcommand{\tib}{T_{\rm i2}}
\newcommand{\tipara}{T_{\rm i\parallel}}
\newcommand{\tiperp}{T_{\rm i\perp}}
\newcommand{\tepara}{T_{\rm e\parallel}}
\newcommand{\teperp}{T_{\rm e\perp}}
\newcommand{\tj}{T_{\rm j}}
\newcommand{\me}{m_{\rm e}}
\newcommand{\mi}{m_{\rm i}}
\newcommand{\fj}{f_{\rm j}}
\newcommand{\fele}{f_{\rm e}}
\newcommand{\fion}{f_{\rm i}}

\newcommand{\ve}{v_{\rm e}}
\newcommand{\vi}{v_{\rm i}}

\newcommand{\mj}{m_{\rm j}}
\newcommand{\chij}{\chi_{\rm j}}
\newcommand{\chie}{\chi_{\rm e}}
\newcommand{\chii}{\chi_{\rm i}}
\newcommand{\zi}{Z_{\rm i}}
\newcommand{\za}{Z_{\rm 1}}
\newcommand{\zb}{Z_{\rm 2}}
\newcommand{\zj}{Z_{\rm j}}
\newcommand{\lmdj}{\lambda_{\rm j}}
\newcommand{\lmde}{\lambda_{\rm e}}
\newcommand{\sumj}{\sum_{\rm j}}

\newcommand{\vth}{v_{\rm th}}
\newcommand{\vthe}{v_{\rm th,e}}
\newcommand{\vthi}{v_{\rm th,i}}
\newcommand{\vthj}{v_{\rm th,j}}
\newcommand{\vtha}{v_{\rm th,1}}
\newcommand{\vthb}{v_{\rm th,2}}
\newcommand{\omgpe}{\omega_{\rm pe}}
\newcommand{\omgpj}{\omega_{\rm pj}}

\newcommand{\vf}{v_{\rm f}}
\newcommand{\vfi}{v_{\rm fi}}
\newcommand{\vfe}{v_{\rm fe}}
\newcommand{\vfj}{v_{\rm fj}}
\newcommand{\vei}{v_{\rm ei}}
\newcommand{\um}{{\rm \mu m}}
\newcommand{\rci}{r_{\rm ci}}
\newcommand{\rce}{r_{\rm ce}}

\newcommand{\vin}{v_{\rm in}}
\newcommand{\vout}{v_{\rm out}}
\newcommand{\rhoin}{\rho_{\rm in}}
\newcommand{\rhoout}{\rho_{\rm out}}
\newcommand{\rhohot}{\rho_{\rm hot}}
\newcommand{\rhocold}{\rho_{\rm cold}}
\newcommand{\rhosingle}{\rho_{\rm single}}

\newcommand{\esin}{S_{\rm in}}
\newcommand{\esout}{S_{\rm out}}
\newcommand{\ekin}{K_{\rm in}}
\newcommand{\ekout}{K_{\rm out}}

\newcommand{\eiout}{u_{\rm out}}
\newcommand{\hin}{H_{\rm in}}
\newcommand{\hout}{H_{\rm out}}

\newcommand{\Bin}{B_{\rm in}}
\newcommand{\Bout}{B_{\rm out}}

\makeatletter
\def\@email#1#2{%
 \endgroup
 \patchcmd{\titleblock@produce}
  {\frontmatter@RRAPformat}
  {\frontmatter@RRAPformat{\produce@RRAP{*#1\href{mailto:#2}{#2}}}\frontmatter@RRAPformat}
  {}{}
}%
\makeatother
\begin{document}

\preprint{AIP/123-QED}

\title{Detection of current-sheet and bipolar ion flows in a self-generated antiparallel magnetic field of laser-produced plasmas for magnetic reconnection research}

\author{T. Morita}
\email{morita@aees.kyushu-u.ac.jp}
\affiliation{Faculty of Engineering Sciences, Kyushu University, 6-1 Kasuga-Koen, Kasuga, Fukuoka 816-8580, Japan}

\author{T. Kojima}
\affiliation{Interdisciplinary Graduate School of Engineering Sciences, Kyushu University, 6-1, Kasuga-Koen, Kasuga, Fukuoka 816-8580, Japan}

\author{S. Matsuo}
\affiliation{Interdisciplinary Graduate School of Engineering Sciences, Kyushu University, 6-1, Kasuga-Koen, Kasuga, Fukuoka 816-8580, Japan}

\author{S. Matsukiyo}
\affiliation{Faculty of Engineering Sciences, Kyushu University, 6-1 Kasuga-Koen, Kasuga, Fukuoka 816-8580, Japan}

\author{S. Isayama}
\affiliation{Faculty of Engineering Sciences, Kyushu University, 6-1 Kasuga-Koen, Kasuga, Fukuoka 816-8580, Japan}

\author{R. Yamazaki}
\affiliation{Department of Physical Sciences, Aoyama Gakuin University, 5-10-1 Fuchinobe, Sagamihara, Kanagawa 252-5258, Japan}

\author{S. J. Tanaka}
\affiliation{Department of Physical Sciences, Aoyama Gakuin University, 5-10-1 Fuchinobe, Sagamihara, Kanagawa 252-5258, Japan}

\author{K. Aihara}
\affiliation{Department of Physical Sciences, Aoyama Gakuin University, 5-10-1 Fuchinobe, Sagamihara, Kanagawa 252-5258, Japan}

\author{Y. Sato}
\affiliation{Department of Physical Sciences, Aoyama Gakuin University, 5-10-1 Fuchinobe, Sagamihara, Kanagawa 252-5258, Japan}

\author{J. Shiota}
\affiliation{Department of Physical Sciences, Aoyama Gakuin University, 5-10-1 Fuchinobe, Sagamihara, Kanagawa 252-5258, Japan}


\author{Y. Pan}
\affiliation{Interdisciplinary Graduate School of Engineering Sciences, Kyushu University, 6-1, Kasuga-Koen, Kasuga, Fukuoka 816-8580, Japan}

\author{K. Tomita}
\affiliation{Faculty of Engineering, Hokkaido University, Kita 13 Nishi 8, Kita-ku, Sapporo 060-8628, Japan}

\author{T. Takezaki}
\affiliation{Faculty of Engineering, University of Toyama, Gofuku 3190, Toyama-shi, Toyama 930-8555, Japan}

\author{Y. Kuramitsu}
\affiliation{Graduate School of Engineering, Osaka University, 2-1 Yamadaoka, Suita, Osaka 565-0871, Japan}

\author{K. Sakai}
\affiliation{Graduate School of Engineering, Osaka University, 2-1 Yamadaoka, Suita, Osaka 565-0871, Japan}

\author{S. Egashira}
\affiliation{Graduate School of Science, Osaka University, 1-1 Machikane-yama, Toyonaka, Osaka 560-0043, Japan}

\author{H. Ishihara}
\affiliation{Graduate School of Science, Osaka University, 1-1 Machikane-yama, Toyonaka, Osaka 560-0043, Japan}

\author{O. Kuramoto}
\affiliation{Graduate School of Science, Osaka University, 1-1 Machikane-yama, Toyonaka, Osaka 560-0043, Japan}

\author{Y. Matsumoto}
\affiliation{Graduate School of Science, Osaka University, 1-1 Machikane-yama, Toyonaka, Osaka 560-0043, Japan}

\author{K. Maeda}
\affiliation{Graduate School of Science, Osaka University, 1-1 Machikane-yama, Toyonaka, Osaka 560-0043, Japan}

\author{Y. Sakawa}
\affiliation{Institute of Laser Engineering, Osaka University, 2-6 Yamadaoka, Suita, Osaka 565-0871, Japan}


\date{\today}

\begin{abstract}
	Magnetic reconnection in laser-produced magnetized plasma is investigated
	by using optical diagnostics.
	The magnetic field is generated via Biermann battery effect,
	and the inversely directed magnetic field lines interact with each other.
	It is shown by self-emission measurement that 
	two colliding plasmas stagnate on a mid-plane forming 
	two planar dense regions, and that they interact later in time.
	Laser Thomson scattering spectra are distorted in the direction of the 
	self-generated magnetic field, 
	indicating asymmetric ion velocity distribution and
	plasma acceleration. In addition, the spectra perpendicular to 
	the magnetic field show different peak intensity, suggesting
	an electron current formation.
	These results are interpreted as magnetic field dissipation, 
	reconnection, and outflow acceleration.
	Two-directional laser Thomson scattering is, as discussed here,
	a powerful tool for 
	the investigation of microphysics in the reconnection region.
\end{abstract}

\maketitle

\section{Introduction}
%
Magnetic reconnection in collisionless plasma plays a key role 
in global change of magnetic field topology and rapid energy conversion 
from magnetic field to plasma thermal and 
kinetic energies\cite{Yamada2010-jd,Zweibel2009-ci}. 
The reconnection physics includes both microscopic magnetic-field
dissipation in an electron scale, 
and macroscopic field advection in surrounding plasmas.
This large-scale difference makes it difficult to understand 
the whole story of magnetic reconnection. 
Laser-plasma experiment can be a useful tool for investigating 
magnetic reconnection, especially in high-beta 
condition.
Strong magnetic field is, spontaneously, generated 
in a high-temperature and high-speed expanding plasma 
via laser-solid interaction, and an anti-parallel field structure
is easily formed by the laser irradiation of two different 
spots\cite{Rosenberg2015-cj,Zhong2010-ko,Nilson2006-aj,Li2007-hy}.
However, local measurements of plasma parameters and magnetic
field are difficult in such small-scale and fast expanding plasmas,
and, so far, there have been few discussions on current sheet formation, 
inflow and outflow parameters, plasma energization, and reconnection rate.

Recently, laser-produced plasmas have been precisely measured with
laser Thomson scattering 
(LTS)\cite{Yamazaki2022-ce,Morita2020-yv,Morita2019-lx,Nilson2006-aj}.
The spectral shape of the ion-feature 
is explained as a result of ion-acoustic resonance
and Landau damping on an ion-acoustic wave depending on 
ion and electron velocity distributions.
Typical plasma parameters such as temperature, density, average charge state,
and flow velocity can be obtained in the case of Maxwellian velocity
distribution.
However, the velocity distributions can be asymmetric in non-equilibrium plasma,
such as in a shock transition region, current-sheet,
and magnetic reconnection region.
Even when ions are in non-Maxwellian, 
the ion distribution function is inferred 
considering the different damping effects on
positive and negative phase velocities, or blue- and red-shifted
resonance peaks of the scattered spectrum.

In this paper, we report the measurement of appearance and disappearance
of an electron current sheet accompanied by bidirectional ion outflows,
for the first time,
in the time-evolution of magnetic reconnection occurring
between laser-produced magnetized plasmas.
The self-emission (SE) imaging shows the interaction of two laser-produced plasmas.
Two plasmas stagnate in an anti-parallel self-generated magnetic field
and they connect with each other later in time,
suggesting sudden decrease in the magnetic pressure.
The resonant peaks of the ion-feature almost perpendicular to 
the self-generated magnetic field, $\vector{B}$,
show 
asymmetry in height, 
suggesting electron drift
relative to ions or asymmetric electron velocity distribution,
in other words, an electron current formation.
The asymmetry in the spectrum decreases later in time,
which means symmetric velocity distribution 
on both electrons and ions. 
This fact indicates the disappearance of the electron current.
The ion-feature parallel to $\vector{B}$ shows different widths
on blue- and red-shifted peaks,
indicating asymmetric ion velocity distribution
or bidirectional ion flows depending on the position.
The appearance and disappearance of the electron current 
and bipolar ion flows are interpreted as 
the magnetic-field dissipation in the current sheet, magnetic reconnection,
and resultant outflow jets.

In the section \ref{sec:TS}, we briefly review the theory of LTS
with Maxwellian and non-Maxwellian electron and ion
velocity distributions.
The experimental observation of asymmetric ion-features, and
interpretations of these spectra with non-Maxwellian velocity
distributions are shown in the section \ref{sec:exp}.
In addition, we discuss the existence of bipolar ion flows 
and electron current in the anti-parallel magnetic field
in the section \ref{sec:discuss},
and we summarize the analysis and discussion in the section \ref{sec:sum}.

\section{Laser Thomson scattering in the case of non-Maxwellian electron and ion distributions}
\label{sec:TS}

\subsection{LTS spectrum for Maxwellian velocity distribution}
\label{sec:tsmax}

LTS spectrum is expressed with the spectral density function\cite{Sheffield2010-hg}:
\begin{eqnarray}
	S(k,\omega) &=& \frac{2{\pi}}{k}\left[ \left| 1-\frac{\chie}{\epsilon} \right|^2\fele(\frac{\omega}{k}) \right. \nonumber \\
	&& \left. + \sumj \frac{\zj^2\NJ}{\NE}\left| \frac{\chie}{\epsilon} \right|^2 \fj(\frac{\omega}{k}) \right],
\end{eqnarray}
where j is the ion species, $\vector{k}=\vector{k}_s-\vector{k}_i$ and 
$\omega=\omega_s - \omega_i$ 
are the wavenumber and frequency of the plasma wave,
respectively, 
$\vector{k}_i$ and $\vector{k}_s$ are the wavenumbers of 
incident and scattered light, respectively,
$\omega_i$ and $\omega_s$ are the frequencies of incident and scattered light,
respectively, and
$\epsilon$ and $\chie$ are 
longitudinal dielectric function and electron susceptibility, respectively, 
shown below:
\begin{eqnarray}
\epsilon &=& 1 + \chie + \sumj \chij,\\
\chie &=& \frac{e^2\NE}{\me \epsilon_0 k} \int\frac{\partial \fele/\partial v}{\omega - kv} dv,\\
\chij &=& \frac{\zj^2 e^2\NJ}{\mj \epsilon_0 k} \int\frac{\partial \fj/\partial v}{\omega - kv} dv.
\end{eqnarray}
$\zj$ is the average charge state of ions, and
$\fele$ and $\fj$ are the electron and ion velocity distributions, respectively.
In the case of Maxwellian, $f$ is expressed below:
\begin{eqnarray}
	f(v) = \frac{1}{\sqrt{\pi}\vth}\exp\left(-\frac{(v-\vf)^2}{\vth^2}\right).
\end{eqnarray}
Here, $\vf$ is the flow velocity in $\vector{k}$ direction,
and $\vth$ is the thermal velocity of species:
\begin{eqnarray}
	\begin{array}{lll}
		\vthe = \sqrt{\frac{2\te}{\me}} & {\rm and} &
\vthj = \sqrt{\frac{2\tj}{\mj}}.
	\end{array}
\end{eqnarray}
When both the ions and electrons are in Maxwellian, $S(k,\omega)$ becomes
\begin{eqnarray}
	S(k,\omega) = \frac{2\sqrt{\pi}}{k}\left[ \frac{1}{\vthe}\left| 1-\frac{\chie}{\epsilon} \right|^2e^{-\XE^2}\right.\nonumber \\
	\left.+ \sumj \frac{1}{\vthj}\frac{\zj^2\NJ}{\NE}\left| \frac{\chie}{\epsilon} \right|^2 e^{-\XJ^2} \right],
\end{eqnarray}
where $\XE$ and $\XJ$ are phase velocities normalized by the thermal velocities
shown below:
\begin{eqnarray}
	\begin{array}{lll}
		\XE = (\omega/k-\vfe)/{\vthe} & {\rm and} & \XJ = (\omega/k-\vfj)/{\vthj}.
	\end{array}
\end{eqnarray}
$\chie$ and $\chij$ become
\begin{eqnarray}
	\begin{array}{lll}
		\chie = \frac{1}{k^2\lmde^2} w(\XJ) & {\rm and} & \chij = \frac{1}{k^2\lmdj^2} w(\XJ),
	\end{array}
\end{eqnarray}
where 
\begin{eqnarray}
	\lmde^2 &=& \vthe^2/\omgpe^2 = 2\epsilon_0\te/\NE e^2,\\
	\lmdj^2 &=& \vthj^2/\omgpj^2 = 2\epsilon_0\tj/\NJ\zj^2e^2,
\end{eqnarray}
and $w$ is the derivative of plasma dispersion function.
In the case of single ion species, we have
\begin{eqnarray}
	S(k,\omega) = \frac{2{\pi}}{k}\left[ \left| 1-\frac{\chie}{\epsilon} \right|^2\fele(\frac{\omega}{k}) + \zi\left| \frac{\chie}{\epsilon} \right|^2 \fion(\frac{\omega}{k}) \right],
\end{eqnarray}
or in Maxwellian,
\begin{eqnarray}
	S(k,\omega) &=& \frac{2\sqrt{\pi}}{k}\left[ \frac{1}{\vthe}\left| 1-\frac{\chie}{\epsilon} \right|^2e^{-\XE^2} \right. \nonumber \\
	&& \left. + \frac{\zi}{\vthi}\left| \frac{\chie}{\epsilon} \right|^2 e^{-\XI^2} \right].\label{eq:sk1-maxwell}
\end{eqnarray}

\subsection{Maxwellian and non-Maxwellian electron velocity distributions}
\label{sec:ts-nomaxe}

\begin{figure}
\begin{center}
\includegraphics[width=0.9\linewidth]{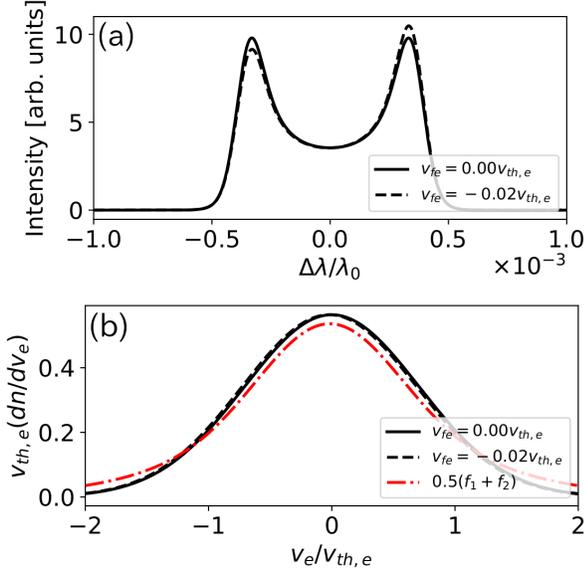}
\caption{\label{fig:TS_ve} 
(a) LTS spectra with two different electron drifts relative to ions:
	$\vfi = 0$, $\vfe=0$ (solid line) and $-0.02\vthe$ (dashed line).
	(b) The electron velocity distributions with the flow velocities
	$\vfe = 0$ (solid line) and $\vfe = -0.02\vthe$ (dashed line).
	A non-Maxwellian electron velocity distribution (dot-dashed line) reproduces
	the same spectrum of $\vfe = -0.02 \vthe$ shown with dashed line in (a).
}
\end{center}
\end{figure}

When the electron flow drifts from ion flow keeping their distributions
in Maxwellian,
LTS spectrum becomes asymmetric due to different rates of
electron and ion Landau damping on the ion acoustic waves
propagating positive and negative $\vector{k}$ directions 
(the left and right sides of the ion-feature).
Figure \ref{fig:TS_ve}(a) shows the ion-features of LTS
with different electron drift $\vfe=0$ (solid line) and $-0.02\vthe$ 
(dashed line). The corresponding velocity distributions are shown
in Fig. \ref{fig:TS_ve}(b) with solid 
and dashed lines, respectively.
This asymmetric LTS spectrum can also be obtained 
when the electron velocity distribution is distorted 
and is no longer in Maxwellian as shown with dot-dashed line 
in Fig. \ref{fig:TS_ve}(b).
Here, the non-Maxwellian distribution is expressed with the summation
of two different Maxwellian distributions:$f = 0.5(f_1+f_2$) with
different temperatures and drift velocities, and
same electron damping effects on the ion-acoustic wave,
indicating the same derivative of the distribution function
$\partial f/\partial v$ near the phase velocity of the ion-acoustic wave.

\subsection{Non-Maxwellian ion velocity distribution}
\label{sec:ts-nomaxi}

\begin{figure}
\begin{center}
\includegraphics[width=0.9\linewidth]{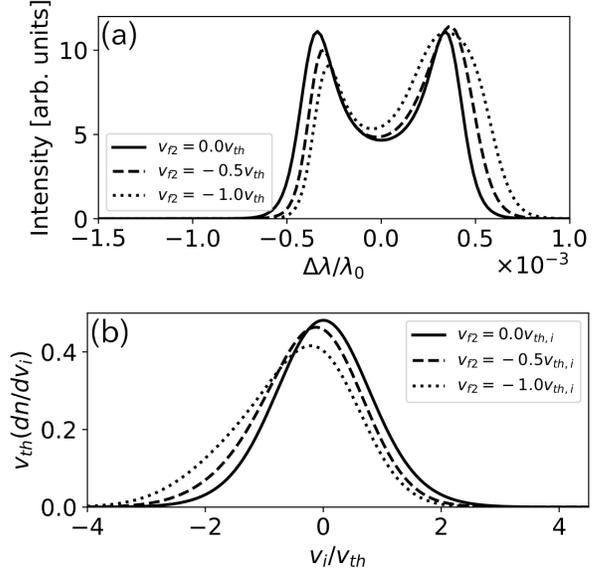}
\caption{\label{fig:TS_vi} 
	(a) LTS spectra for non-Maxwellian ion velocity distributions of
	$f = f_1 + f_2$, where $f_1$ and $f_2$ are Maxwellian distributions
	with different temperatures and drift velocities 
	${\vf}_2 = 0$,
	$-0.5\vthi$, and $-\vthi$
	shown with
	solid, dashed and dotted lines, respectively).
	The corresponding velocity distributions are shown in (b).
}
\end{center}
\end{figure}

When the ion velocity distribution is not expressed with Maxwellian,
the ion susceptibility is not so simple,
and an ion-feature is affected by different Landau damping 
on left and right peaks depending on ion and electron 
velocity distributions, 
which has been experimentally observed\cite{Ross2012-le,Swadling2020-rc}
and numerically calculated\cite{sakai2022} for two-streaming plasmas.
We assume 
that non-Maxwellian ion distribution $\fion$ 
is the sum of two different Maxwellian with different temperatures as follows:
\begin{eqnarray}
	\fion = \alpha f_{\rm i1} + (1-\alpha)f_{\rm i2},
\end{eqnarray}
where $\alpha$ is the abundance ratio. The ion susceptibility $\chii$ becomes
\begin{eqnarray}
	\chii &=& \alpha{\chii}_1 + (1-\alpha){\chii}_2,
\end{eqnarray}
and LTS spectrum $S(k,\omega)$ is given by
\begin{eqnarray}
	S(k,\omega) &=& \frac{2\sqrt{\pi}}{k}\left[ \frac{1}{\vthe}\left| 1-\frac{\chie}{\epsilon} \right|^2e^{-\XE^2} + \frac{\alpha\za}{\vtha}\left| \frac{\chie}{\epsilon} \right|^2 e^{-{\XIA}^2}\right.\nonumber\\
	&+& \left.\frac{(1-\alpha)\zb}{\vthb}\left| \frac{\chie}{\epsilon} \right|^2 e^{-{\XIB}^2} \right], \label{eq:sk2-maxwell}
\end{eqnarray}
as shown in Fig. \ref{fig:TS_vi}(a).
We assume that electrons are in Maxwellian, and that
there is no effective current: 
$\vfe\NE = \alpha Z_1{\NI}_1{\vf}_1 + (1-\alpha)Z_2{\NI}_2{\vf}_2$.
Note that the ion-feature shown in Fig. \ref{fig:TS_vi}(a) does not strongly
depend on the electron flow velocity in $0<|\vfe|\lesssim|{\vf}_2|$.
Figure \ref{fig:TS_vi}(b) shows the ion velocity distributions 
with $\alpha = 0.5$, ${\ti}_1=100$ eV, ${\ti}_2=300$ eV,
\begin{eqnarray}
	{\fion}_1(v) &=& \frac{1}{\sqrt{\pi}\vtha}\exp\left(-\frac{(v-{\vf}_1)^2}{\vtha^2}\right),\\
	{\fion}_2(v) &=& \frac{1}{\sqrt{\pi}\vthb}\exp\left(-\frac{(v-{\vf}_2)^2}{\vthb^2}\right), 
\end{eqnarray}
and the flow velocities of first and second flows: ${\vf}_1 = 0$ 
and ${\vf}_2 \ne 0$.
Here, three velocity distributions with different ${\vf}_2$ 
of 0, $-0.5\vthi$, and $-\vthi$ are shown.

Unlike in the case of electron drift shown in Fig. \ref{fig:TS_ve},
both the intensity and the width of the two resonant peaks change depending on
$\tia$, $\tib$, $\vf$, and $\alpha$,
and this asymmetric effect on each resonant peak allows us to infer
rapid thermalization and acceleration such as 
the interaction of counter-streaming plasmas\cite{Ross2012-le,Ross2013-vs,Park2012-pq,Sakawa2017-dw,Kugland2012-hw}, 
shockwave generation\cite{Fiuza2020-pf,Li2019-jy,Schaeffer2019-ei,Schaeffer2017-ik}, and
magnetic reconnection\cite{Totorica2020-fn,Zhong2018-nd,Raymond2018-np}.

\subsection{Modified spectrum measured with gated detector}

\label{sec:tsvave}

\begin{figure}
\begin{center}
\includegraphics[width=1.0\linewidth]{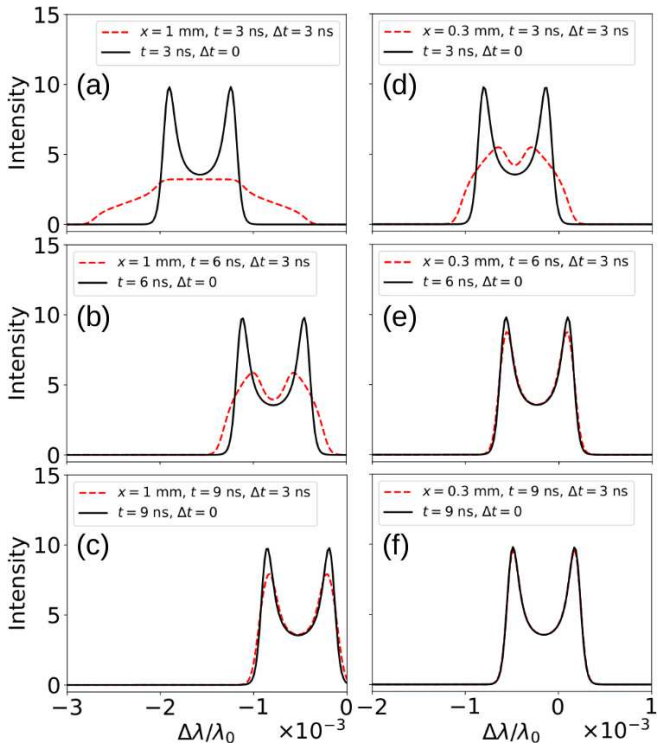}
\caption{\label{fig:TS_vave}
	The ion-features of LTS spectra calculated with Eq. (\ref{eq:sk1-maxwell})
	(solid line) and 
	Eq. (\ref{eq:TSvave}) (dashed line),
	with the measurement time $t=3$ [(a) and (d)], 
	6 [(b) and (e)], and 9 ns [(c) and (f)],
	the exposure time $\Delta t = 3$ ns, 
	and measurement position of 
	$x=1$ mm [(a), (b), and (c)] and
	0.3 mm [(d), (e), and (f)].
	The carbon plasma is assumed with $\te=\ti=100$ eV, 
	$\NE=2\times10^{18}$ cm$^{-3}$, $Z=6$, and 
	$\vf=x/t$.
}
\end{center}
\end{figure}

In the case of experimental measurement,
the spectrum is modified when the velocity changes in an exposure time
of a detector.
This effect is sometimes important for pulsed plasmas,
for example, laser-produced plasmas, measured with a gated detector
such as intensified charge coupled device (ICCD) camera.
The spectrum is modified by taking the average in 
$\vf \pm \Delta v/2$
as
\begin{eqnarray}
	S_{\Delta \vf}(k,\omega) = \frac{1}{\Delta \vf}\int_{\vf-\Delta v/2}^{\vf+\Delta v/2} S(k,\omega)d\vf. \label{eq:TSvave}
\end{eqnarray}
The flow velocity is typically expressed as $\vf = x/t$, 
where $x$ is the distance and $t$ is the time after the laser-irradiation,
and the flow velocity change in a gate width of $\Delta t$
becomes 
	$\Delta \vf \sim \vf\Delta t/t$. 
This modification can be ignored 
	with small $\vf$ and/or $\Delta t/t \ll 1$, but
should be taken into account for fitting the measured spectrum
with large $\vf$ and $\Delta t/t \gtrsim 1$.
Figures \ref{fig:TS_vave}(a)--\ref{fig:TS_vave}(c) show the spectra 
of the ion-features for carbon plasma 
with $\te=\ti=100$ eV, $\NE = 2\times10^{18}$ cm$^{-3}$, $Z=6$
at $x=1$ mm and
at $t=3$, 6, and 9 ns, respectively.
Figures \ref{fig:TS_vave}(d)--\ref{fig:TS_vave}(f) show the spectra 
calculated in the same way with the distance $x=0.3$ mm.
The dashed lines show the velocity-averaged spectra calculated
with Eq. (\ref{eq:TSvave}) with the gate width $\Delta t = 3$ ns
which is comparable to the present experiment shown later.
The velocity-averaged spectra (dashed lines) 
become close to the theoretical spectra ($\Delta t = 0$, solid lines) 
later in time
in the both cases of the distance $x=1$ mm and 0.3 mm.
Even in the case of small $\Delta t/t$ with $x=1$ mm, for example, 
$\Delta t/t = 1/2$ [Fig. \ref{fig:TS_vave}(b)] and 
$1/3$ [Fig. \ref{fig:TS_vave}(c)],
the spectra are modified.
On the other hand, with $x=0.3$ mm, the modification is small
as shown in Figs. \ref{fig:TS_vave}(e) and \ref{fig:TS_vave}(f).
Our spectral analyses in the following sections are done
in the time range of $t=5$--9 ns and in the distance
$-0.3$ mm $< x < 3$ mm,
and this velocity-averaged effect in a limited gate width is ignorable.
Therefore, LTS spectra are fitted with the theoretical function
without taking average in velocity in the following sections.

\section{Experiment}
\label{sec:exp}

\begin{figure}
\begin{center}
\includegraphics[width=1.0\linewidth]{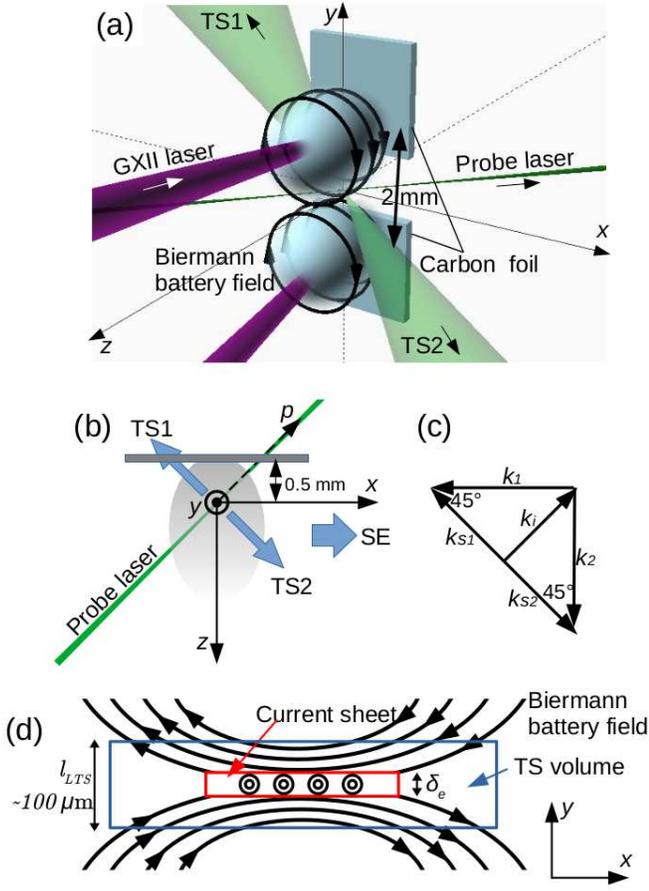}
\caption{\label{fig:setup}
	(a) Schematic view of laser ablation, self-generated (Biermann battery)
	magnetic field, and LTS measurement.
	(b) The top view of the target showing the laser-produced plasma,
	probe laser, and TS and SE measurements.
	(c) Two different TS measurement directions 
	($\vector{k_1}$ and $\vector{k_2}$).
	(d) The magnetic field structure on 
	the $xy$-plane.
}
\end{center}
\end{figure}

\begin{figure}
\begin{center}
\includegraphics[width=0.9\linewidth]{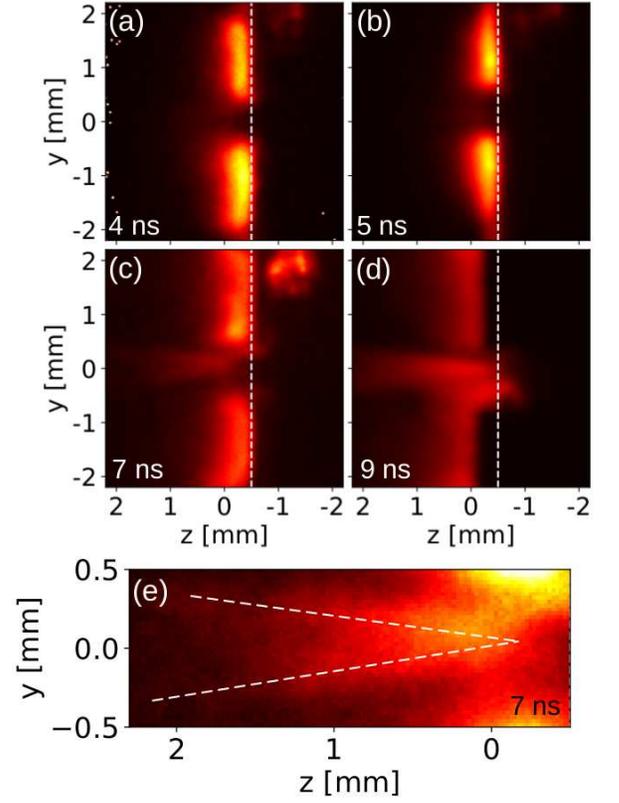}
\caption{\label{fig:se} 
	SE images at (a) $t=4$, (b) 5, (c) 7, and (d) 9 ns.
	(e) The enlarged figure of colliding plasmas at $t=7$ ns (c).
	Two plasmas expanding from the upper and lower targets slow down
	by the piled-up magnetic field near the plane $y=0$,
	and dense structures are formed along the dashed lines.
}
\end{center}
\end{figure}

The experiment was performed with Gekko-XII laser system at
the Institute of Laser Engineering, Osaka University.
Two laser beams with the average energy of 700 J in 1.3 ns at the wavelength
of 1053 nm were focused in the diameter of $\sim$100 $\um$ and
irradiate two individual carbon planar foils with the thickness of 
0.1 mm.
Two foils were located at $z=-0.5$ mm on the $x$-$y$ plane, and
two spots were separated by 2 mm along the $y$-axis 
as shown in Fig. \ref{fig:setup}(a).
Strong magnetic field of $B\sim100$ T is self-generated 
due to the anisotropy of density and temperature gradients  
formed around the laser spots 
(Biermann battery effect\cite{Stamper1975-me},
$\partial \vector{B}/\partial t \propto \vector{\nabla} \te \times \vector{\nabla} \NE$).
As discussed later, laser-produced plasma has large plasma beta
($\beta_e = 2\mu_0\NE\te/B^2\gg1$), indicating that the magnetic field
is advected along with freely expanding electron flux.
$\beta_e$ becomes small later in time, and $\beta_e \lesssim 1$
when two plasmas interact at $t\sim$ 5--7 ns.
Two anti-parallel magnetic-field lines interact at $x\sim0$ on the mid-plane,
$y=0$,
as the plasma plumes expand.
Similar
experimental setup has recently been used for magnetic reconnection
research with high-power laser\cite{Nilson2006-aj,Li2007-hy,Nilson2008-le,Fox2012-wm,Zhong2010-ko,Rosenberg2015-wt,Rosenberg2015-cj},
and the current sheet formation and magnetic reconnection 
have been investigated with particle-in-cell simulations\cite{Matteucci2018-pi,Fox2018-wk}.
Another laser (probe laser, Nd:YAG) with the energy of 330 mJ in 10 ns
at the wavelength of 532 nm focused at the origin $(x,y,z) = (0,0,0)$,
and the Thomson scattered light was detected from two directions
(TS1 and TS2).
The probe laser direction was 45$^\circ$ from $x$ and $z$ axes, and 
the axis $p$ is defined along the probe laser: 
$(x,y,z) = (p/\sqrt{2},0,-p/\sqrt{2})$.
LTS measures local plasma parameters along the probe laser, and
the spatial resolution is determined by the focal spot,
$l_{\rm LTS}\sim$100 $\um$, as shown in Fig. \ref{fig:setup}(d).
The resolution in wavelength is determined by the entrance slit and
the dispersion in the spectrometer, which is directly measured by observing 
Rayleigh scattering from nitrogen gas filled in the vacuum chamber.
We used high wavelength-resolution spectrometer with triple-grating
systems\cite{Tomita2017-jr,Morita2019-lx,Yamazaki2022-ce} 
and the resolutions were $25\pm1$ pm for TS1
and $20\pm1$ pm for TS2,
and the dispersed light was detected 
with ICCD cameras with the gate widths of 3 ns.
The top view of this geometry is also illustrated 
in Fig. \ref{fig:setup}(b).
These two diagnostics measure plasma parameters
in two different directions ($\vector{k_1}$ and $\vector{k_2}$) as shown 
in Fig. \ref{fig:setup}(c).
Plasma density structure was also imaged with an ICCD camera 
with the gate width of 0.2 ns
by observing a self-emission
at the wavelength of $450\pm5$ nm.

\subsection{Self-emission imaging}

Figures \ref{fig:se}(a)--\ref{fig:se}(d) show the SE images
taken from $t=4$ to 9 ns.
The dashed-lines show the surface of the carbon target.
Generally, the emission intensity is interpreted as thermal bremsstrahlung
emission in optically thin plasma, 
and it strongly depends on the electron density\cite{Rybicki1980-my}.
Early in time at $t=4$--5 ns [Figs. \ref{fig:se}(a) and \ref{fig:se}(b)], 
two plasmas begin to interact on the mid-plane, $y=0$ mm.
Two planar structures are formed 
at $y>0$ mm and $y<0$ mm at $t=7$ ns as shown in Fig. \ref{fig:se}(c).
As the two plasmas with self-generated magnetic fields expand,
anti-parallel field structures would be piled-up on the 
mid-plane\cite{Matteucci2018-pi,Li2019-jy,Nilson2008-le,Fox2012-wm,Li2016-zx},
and 
the
stronger field decelerate the plasma expansion, forming these
dense structures as shown with dashed-lines in the enlarged figure
[Fig. \ref{fig:se}(e)].
These dense structures begin to merge at $z\sim0$
at $t=7$ ns,
and continue to merge at $z>0$ later in time
forming a single planar structure as shown in Fig. \ref{fig:se}(g).
This indicates the plasma stagnation due to the decrease in magnetic pressure 
on the mid-plane.

\subsection{LTS parallel to \vector{B}}

\begin{figure}
\begin{center}
\includegraphics[width=1.0\linewidth]{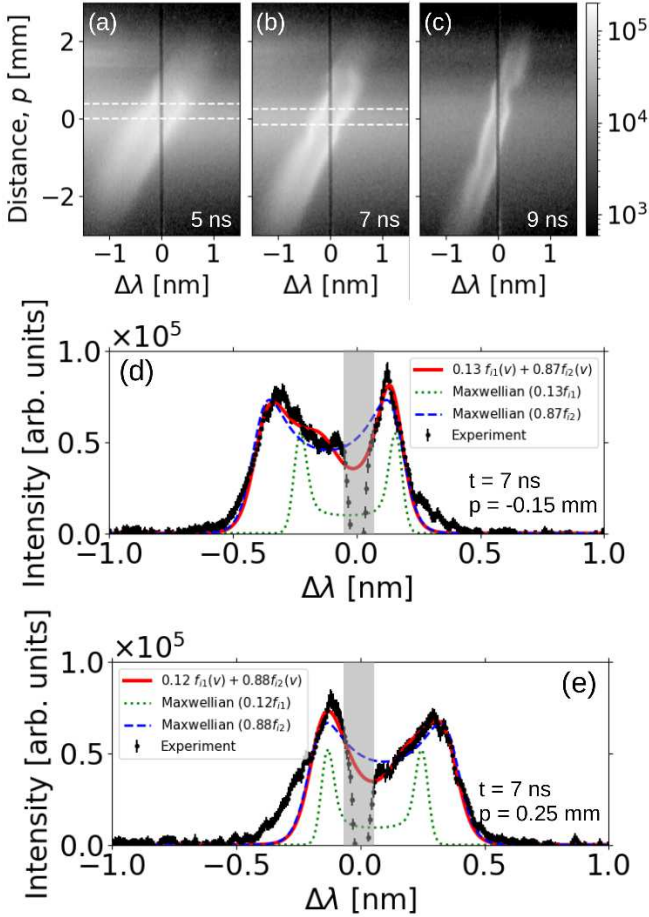}
\caption{\label{fig:TS_para} 
	LTS spectra along $\vector{k_1}$ at (a) $t=5$, (b) 7,
	and (c) 9 ns. (d), (e) 
	The line-out plots at $t=7$ ns
	at (d) $p=-0.15$ mm and (e) $0.25$ mm.
	These spectra are fitted with Eq. (\ref{eq:sk2-maxwell}) and shown
	with solid lines. Theoretical functions with single Maxwellian
	are shown with dotted ($f_1$) and dashed ($f_2$) lines.
	The shaded areas are affected by a notch filter or stray light.
}
\end{center}
\end{figure}
\begin{figure}
\begin{center}
\includegraphics[width=1.00\linewidth]{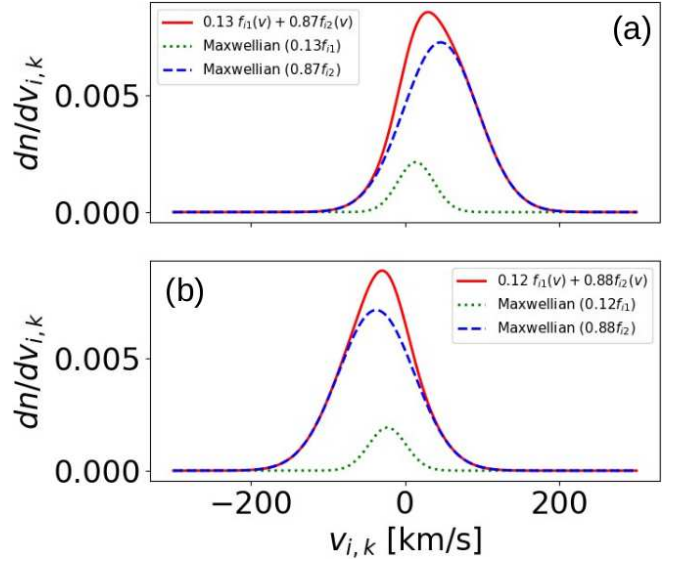}
\caption{\label{fig:TS_para_vi}
	Ion velocity distributions with the parameters obtained from the fitting
	of Figs. \ref{fig:TS_para}(d) and \ref{fig:TS_para}(e).
}
\end{center}
\end{figure}
\begin{figure}
\begin{center}
\includegraphics[width=0.9\linewidth]{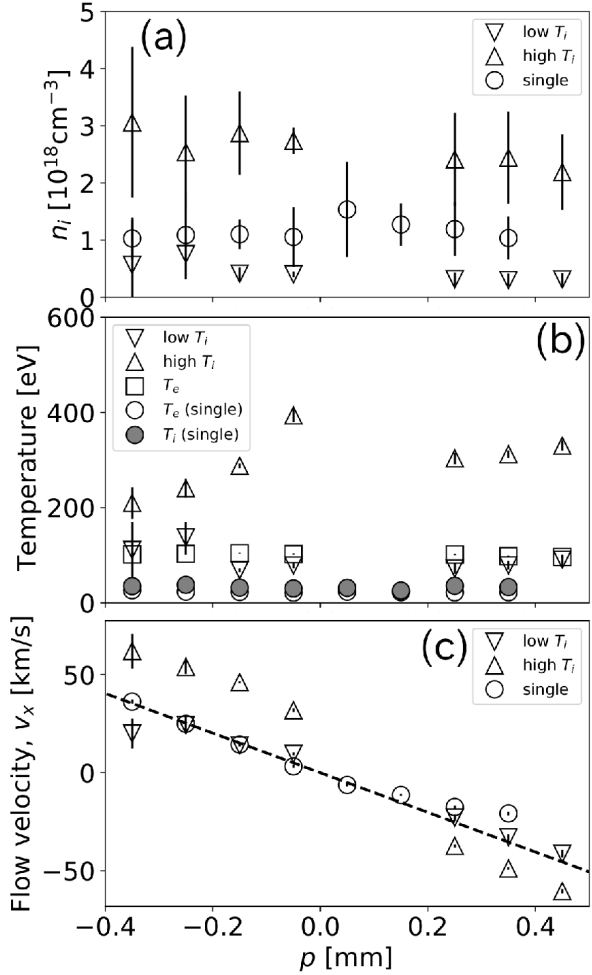}
\caption{\label{fig:TS_para_param}
	(a) Ion density, (b) electron and ion temperatures,
	and (c) flow velocities obtained from the fitting of LTS spectra at
	$t=7$ ns. 
	The plasma parameters obtained from a single laser shot are
	also shown with circles.
	The velocity calculated from the distance $p$ and time $t$,
	$v = p/\sqrt{2}t$, is plotted as a reference for free-streaming velocity.
}
\end{center}
\end{figure}

LTS spectra parallel to the $\vector{k_1}$ vector 
[see Fig. \ref{fig:setup}(c)]
measured in the direction of TS1
at $t=5$, 7, and 9 ns
are shown in Figs. \ref{fig:TS_para}(a)--\ref{fig:TS_para}(c),
respectively.
The observed two peaks show the ion-feature,
and the peak separation is proportional to the sound velocity
$\omega/k \sim c_s \sim [(Z\te+3\ti)/\mi]^{1/2}$.
The widths of the spectra decrease from $t=5$ ns to 9 ns,
suggesting that the temperature decreases in time.
Small fluctuations in bulk-flow velocity are observed at $t=5$ and 7 ns 
around $p\sim0$ mm 
[dashed lines in Figs. \ref{fig:TS_para}(a) and \ref{fig:TS_para}(b)],
and different widths in left and right peaks are seen as well.
This difference is easily seen in the line-out plots shown 
in Figs. \ref{fig:TS_para}(d) and \ref{fig:TS_para}(e),
at $p=-0.15$ and 0.25 mm, respectively.
The light intensity around $\Delta\lambda\sim0$ decreases by a notch filter
as shown in the shaded areas.
These spectra show different widths for the left and right peaks and
this difference can not be interpreted using Maxwellian distributions
for electrons and ions, but
explained with non-Maxwellian ion velocity distribution as explained
in the section \ref{sec:ts-nomaxi}.
Assuming the ion velocity distribution as 
a superposition of two Maxwellians,
observed asymmetric features can be expressed.
Here, collisional-radiative model is assumed and the average
charge state $\zi$ is evaluated as a function of $\te$ and $\NE$,
$\zi = \zi(\te, \NE)$,
using FLYCHK code\cite{H-K_Chung2005-ep}.
The solid-lines are the best-fit results, and 
the dotted and dashed lines are the spectra calculated from
Maxwellian ion velocity distributions.
Here, we assume that the plasma is in the collisionless regime 
for LTS measurement, that is,
the ion-ion mean free path for thermal ions, $\lambda_{\rm i}$,
is much larger than $1/k$, or $k\lambda_{\rm i} \gg 1$,
where $k$ is the wave number of ion-acoustic wave\cite{Myatt1998-el}.
For example, 
$\lambda_{\rm i} \sim 12\pi^2\epsilon_0^2(k_{\rm B}\ti)^2/\sqrt{\pi}Z^4e^4\NI\ln\Lambda = 4.3$ $\um$\cite{Myatt1998-el}
where $\ln\Lambda \sim 6.2$ is the Coulomb logarithm,
with the typical parameters of 
$\ti = 300$ eV, $Z = 4$, and $\NI = 2.7\times10^{18}$ cm$^{-3}$,
$\NE = 1.7\times10^{19}$ cm$^{-3}$,
and $k= k_1 = k_2 = 1.7\times10^7$ m$^{-1}$, resulting in $k\lambda_{\rm i}\sim 72$,
and the collisional effect is small in the present experiment.

The corresponding velocity distributions are shown 
in Figs. \ref{fig:TS_para_vi}(a) and \ref{fig:TS_para_vi}(b).
The best-fit results (solid-lines) show a non-Maxwellian distribution
consisting of two ion distributions with different temperatures and 
drift velocities (dashed and dotted lines).
When the ions drift in the $\vector{k_1}$ direction at $p=-0.15$ mm
[Fig. \ref{fig:TS_para}(d)],
the observed spectrum is explained with 
$0.13f_1 + 0.87f_2$, where
$f_1$ and $f_2$ are Maxwellian distributions
with temperatures of $69\pm4$ eV and $290\pm5$ eV, respectively,
and drift velocities of $14\pm1$ km/s and $46\pm1$ km/s, respectively.
On the other hand, ions drift in the $-\vector{k_1}$ direction at $p=0.25$ mm,
and expressed with 
$0.12f_1+0.88f_2$,
where
$f_1$ and $f_2$ are Maxwellian distributions with
$\ti=72\pm11$ eV and $\vi=-23\pm2$ km/s, 
and $\ti=300\pm12$ eV and $\vi=-37\pm1$ km/s, respectively.

The ion density, electron and ion temperatures, and flow velocity 
at $t=7$ ns
as a function of position $p$ are obtained from the fitting of 
the spectrum of Fig. \ref{fig:TS_para}(b)
and shown in Fig. \ref{fig:TS_para_param}.
The observed spectrum at 
$-0.4$ mm $<p<0.4$ mm are well fitted
with two components with low--density and low--temperature ions
(inverted triangles)
and high--density and high-temperature ions (triangles)
shown in Figs. \ref{fig:TS_para_param}(a) and \ref{fig:TS_para_param}(b).
The electron temperatures [squares in Fig. \ref{fig:TS_para_param}(b)] 
are almost same as the ion temperatures of the lower-temperature component.
In addition, the lower-temperature ions show slower flow-velocity
and this flow velocity is comparable to a free-streaming velocity
calculated as $v = p/\sqrt{2}t$
as shown with dashed line 
in Fig. \ref{fig:TS_para_param}(c).
On the other hand, the ion population with higher-velocity 
and higher-temperature
are observed as shown with triangles in Fig. \ref{fig:TS_para_param}(c).
This velocity shift from the free-streaming velocity 
($\Delta v \sim 30$ km/s) can be interpreted as an acceleration.
These figures suggest that about 90\% of the ions are thermalized 
and accelerated in the $\pm \vector{k_1}$ directions
(or $\pm x$ directions)
which are consistent with the directions of outflows from
a magnetic reconnection between the anti-parallel
self-generated Biermann battery fields.

We also performed a laser shot with only a single beam for plasma generation,
and the ion density, temperatures, and flow velocity are shown with circles
in Figs. \ref{fig:TS_para_param}(a)--\ref{fig:TS_para_param}(c),
respectively.
The flow velocity in the case of single laser shot is almost same as 
the free-streaming velocity shown with dashed line 
in Fig. \ref{fig:TS_para_param}(c), and similar to slower ion component
(inverted triangles).
Therefore, the slower (lower temperature) ion component is considered 
as the plasma directly expands from the laser-spot, while
the faster (higher temperature) plasma is interpreted as the outflow
energized by a magnetic reconnection.

\subsection{LTS perpendicular to \vector{B}}

\begin{figure}
\begin{center}
\includegraphics[width=1.0\linewidth]{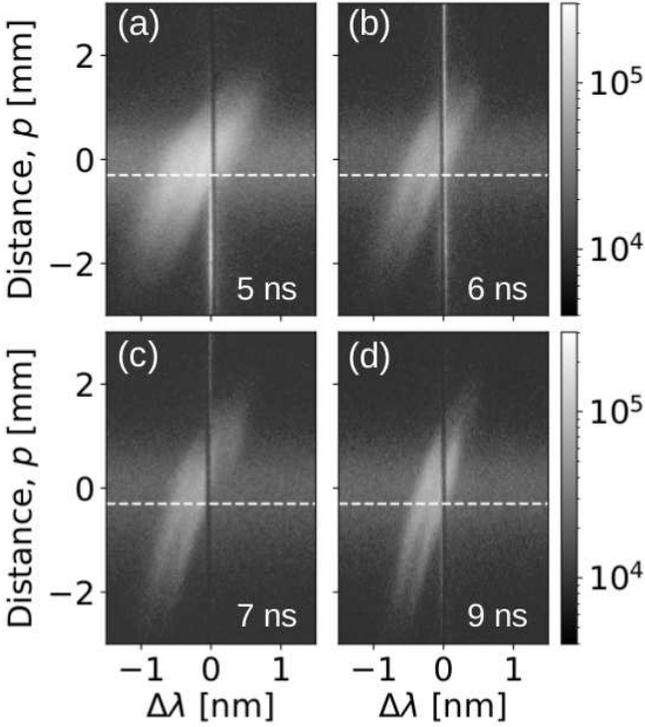}
\caption{\label{fig:TS_perp} 
	LTS spectra with $\vector{k} \perp \vector{B}$ at (a) $t=5$, (b) 6,
	(c) 7, and (d) 9 ns.
}
\end{center}
\end{figure}
\begin{figure}
\begin{center}
\includegraphics[width=0.9\linewidth]{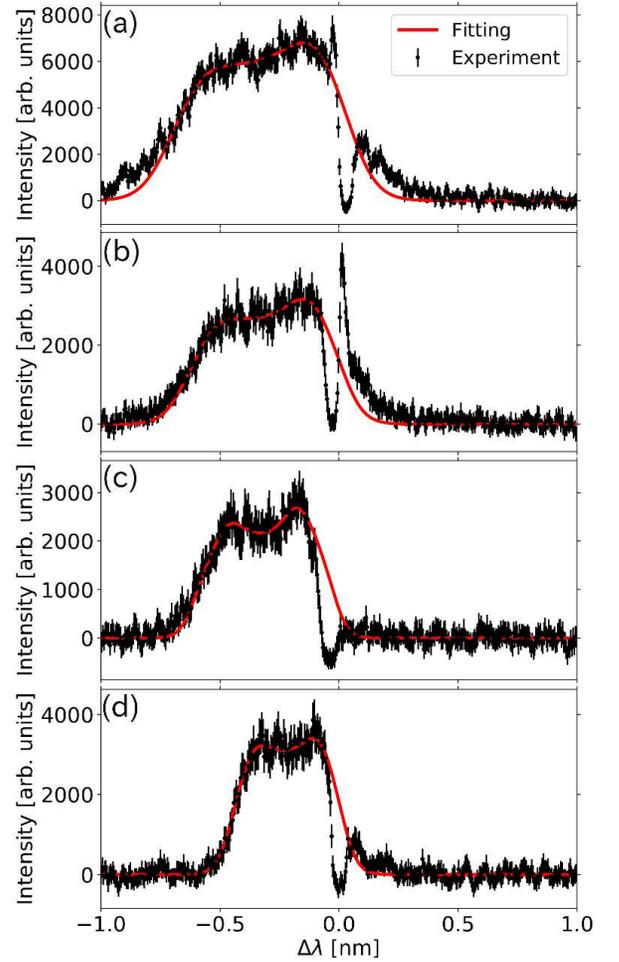}
\caption{\label{fig:TS_perp_spec} 
	LTS spectra at $p=-0.3$ mm with $k \perp \vector{B}$ at 
	(a) $t=5$, (b) 6, (c) 7, and (d) 9 ns.
}
\end{center}
\end{figure}
\begin{figure}
\begin{center}
\includegraphics[width=1.0\linewidth]{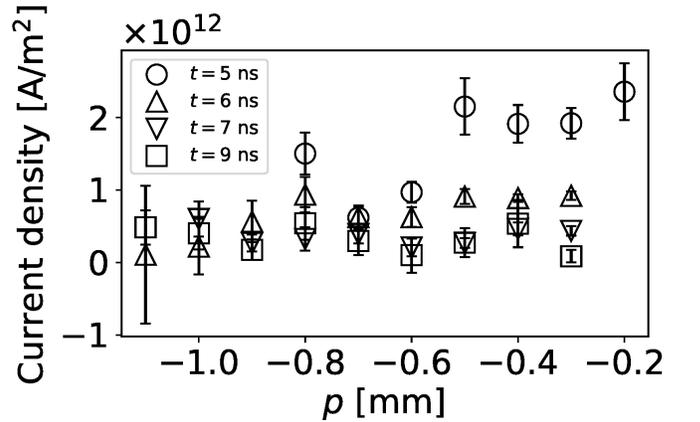}
\caption{\label{fig:current} 
	The current density calculated at $t=5$, 6, 7, and 9 ns
	as a function of position $p$, assuming that electrons and ions
	are in Maxwellian velocity distributions.
}
\end{center}
\end{figure}

Figures \ref{fig:TS_perp}(a)--\ref{fig:TS_perp}(d) show LTS spectra
along $\vector{k_2}$ (perpendicular to the initial anti-parallel magnetic field $\vector{B}$)
at $t=5$, 6, 7, and 9 ns, respectively.
Unlike the spectra in the $\vector{k_1}$ direction (see Fig. \ref{fig:TS_para}),
the spectra are almost straight as a function of position $p$
with constant width, meaning 
no characteristic change in velocity nor temperature.

Figures \ref{fig:TS_perp_spec}(a)--\ref{fig:TS_perp_spec}(d) show 
the line-outs of the Figs. \ref{fig:TS_perp}(a)--\ref{fig:TS_perp}(d)
at $p=-0.3$ mm at $t=5$, 6, 7, and 9 ns, respectively.
The right-peaks are stronger than the left-peaks from $t=5$ to 7 ns.
However, the asymmetry is getting weaker as time evolves, and
at 9 ns, the double-peak becomes almost symmetric. 
This asymmetry is often interpreted as the different
Landau damping on the ion-acoustic waves 
in the $\pm \vector{k_2}$ directions (section \ref{sec:ts-nomaxe}).
When the ions and electrons are in Maxwellian velocity distribution,
this difference occurs with different drift velocities of
electron and ion flows.
The solid lines show the results of fitting assuming
Maxwellian distributions for both ions and electrons,
resulting in
$\te\sim96$, 77, 68, and 45 eV,
$\ti\sim390$, 310, 200, and 220 eV,
$\vei\sim650$, 430, 270, and 100 km/s, 
$\NE\sim2.1\times10^{19}$, $1.3\times10^{19}$, 
$9.9\times10^{18}$, and $1.3\times10^{19}$ cm$^{-3}$, 
and therefore, the estimated current density,
$j_z = Ze\NI\vi-e\NE\ve\sim1.9\times10^{12}$, $9.2\times10^{11}$, $4.4\times10^{11}$,
and $8.7\times10^{10}$ Am$^{-2}$,
respectively,
at $t=5$, 6, 7, and 9 ns.
Figure \ref{fig:current} shows the current density, $j_z$,
as a function of position $p$ at $t=5$, 6, 7, and 9 ns,
where we assume both electrons and ions are in Maxwellian.
Although the spectra are affected by a notch filter at $\Delta\lambda\sim0$
and those at $p>0$ could not be analyzed,
the current is detected in the region $-0.5$ mm $<p<0$ mm,
suggesting that the anti-parallel field structure formation
near $y\sim0$.
$j_z\sim2\times10^{12}$ Am$^{-2}$ at $t=5$ ns is the largest, and
it decreases and almost disappears at $t=9$ ns.



\section{Discussion}
\label{sec:discuss}

\begin{figure}
\begin{center}
\includegraphics[width=1.0\linewidth]{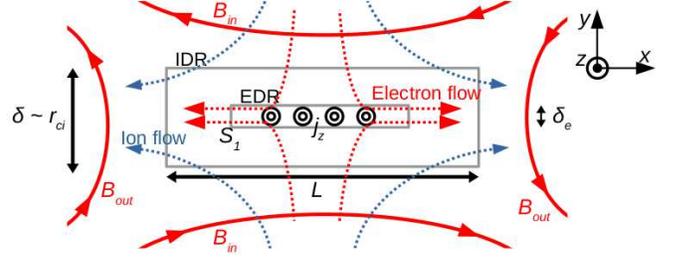}
\caption{\label{fig:MRregion} 
	Structure of magnetic reconnection layer.
	Ions are magnetized outside of the IDR (the region $L\times\delta$) 
	and electrons are still magnetized.
	Electron current is formed and magnetic field diffuses in EDR
	(in the thickness of $\delta_e$).
}
\end{center}
\end{figure}

As previously reported in many researches
with laser-produced plasmas\cite{Matteucci2018-pi,Li2019-jy,Nilson2008-le,Fox2012-wm,Li2016-zx},
a stable magnetic field structure with anti-parallel directions
is formed between two laser-produced plasma plumes
as shown in Figs. \ref{fig:setup}(a), \ref{fig:setup}(d), 
and \ref{fig:MRregion}.
In such a field structure, an electron current would be generated
in an electron dissipation region (EDR) satisfying
\begin{eqnarray}
	\int_{S_1} \mu_0\vector{j}_z\cdot \vector{dS_1} = \int_{S_1} \left( \vector{\nabla} \times \vector{B} \right) \cdot \vector{dS_1} = \oint_{\delta S_1} \vector{B} \cdot \vector{dl},
\end{eqnarray}
where $\vector{j}_z$ is the electron current and
$S_1$ is the cross section of the current sheet.
This current $\vector{j}_z$ flows perpendicular to the plane 
on the anti-parallel magnetic field 
($+z$ or $+\vector{k_2}$ direction 
in Figs. \ref{fig:setup} and \ref{fig:MRregion}).
Taking the thickness of the current sheet as $\delta_e$,
the magnetic field strength $\Bin$ in the upstream region is
\begin{eqnarray}
	\Bin \sim \frac{\mu_0}{2}\delta_e j_z. \label{eq:Binj}
\end{eqnarray}

In general, the thickness of meandering charged particles is estimated as
$d \sim \sqrt{r_{\rm c}\lambda_B}$,
where $r_{\rm c}$ is the Larmor radius for the magnetic field of $\Bin$,
assuming the magnetic field strength $B_x(y) = -\Bin y/\lambda_B$ 
in a plasma sheet\cite{Hoshino2018-wg}.
Here, the magnetic field varies in $\lambda_B$
which is comparable to an ion diffusion region (IDR), $\delta$, 
where ions are unmagnetized and electrons are still magnetized. 
This thickness $\delta$ ($\sim \lambda_B$) should be comparable 
to ion meandering thickness $d$,
and therefore, $d \sim r_{ci} \sim \delta \sim \lambda_B$.
For electrons, the thickness of electron meandering or 
that of electron current sheet is 
\begin{eqnarray}
	\delta_e \sim \sqrt{\rce\lambda_B}\sim \sqrt{\rce\rci}.\label{eq:jB}
\end{eqnarray}

As shown in Figs. \ref{fig:TS_perp_spec}(a)--\ref{fig:TS_perp_spec}(d),
the electrons drift relative to ions from $t=5$ to 9 ns
in the $\vector{k_2}$ direction, suggesting the electron current formation
in the anti-parallel magnetic field.
Also this asymmetry decreases at $t=9$ ns, meaning the disappearance
of this electron current.
However, as explained in the section \ref{sec:ts-nomaxe},
this asymmetric spectrum can be expressed not only by Maxwellian
electrons and ions with different flow velocities
but also by non-Maxwellian electron velocity distribution
with different Landau damping rates on the ion-acoustic waves in 
the $\pm \vector{k_2}$ directions,
indicating that the electron flow velocity and current density are not
constrained only from this asymmetric ion-feature.
Therefore, we estimate $j_z$ and $\Bin$ 
not from TS2 ($\vector{k_2}$ or $z$ direction) 
but from TS1 ($\vector{k_1}$ or $x$ direction) assuming that 
the acceleration in the $\pm x$ directions 
are caused by magnetic reconnection and the outflow
velocity is determined by the Alfven velocity.

Here, considering the conservations of mass and energy during
a magnetic reconnection:
\begin{gather}
	\rhoin L \vin = \rhoout \delta \vout, \label{eq:mass} \\
	\left(\esin + \ekin + \hin \right)\vin L \nonumber \\
	 = \left(\esout + \ekout + \hout \right)\vout \delta,
\end{gather}
where $Sv=(B^2/\mu_0)v$ is Poynting flux, 
$Kv=(\rho v^2/2)v$ and $Hv=(u+p)v$ are
kinetic and enthalpy fluxes, respectively,
$\rho=\mi\NI$ is the mass density,
$u=p/(\gamma-1)$ is the internal energy,
$p=nT$,
and $L$ and $\delta$ represent the length and width of 
the IDR shown in Fig. \ref{fig:MRregion}
with the thickness of $\delta \sim \rci$.
As suggested in Figs. \ref{fig:TS_para_param}(b) and \ref{fig:TS_para_param}(c),
the kinetic energy is $\ekout/\NI \sim \mi\vout^2/2\sim60$ eV using
the accelerated velocity of $\sim30$ km/s,
while the internal energy is 
$\eiout/\NI\sim300/(\gamma-1)$ eV,
indicating $\eiout \sim 5\ekout/(\gamma-1)$.
When the Poynting flux 
and kinetic energy density flux
of the inflow are
converted to 
the kinetic and enthalpy fluxes in the outflow
(assuming $\hin = \esout=0$),
the energy equation becomes
\begin{eqnarray}
	\left(\frac{\Bin^2}{\mu_0} + \frac{\rhoin \vin^2}{2} \right)L\vin \sim \frac{6\gamma-1}{\gamma-1}\ekout \delta\vout.
\end{eqnarray}
By using Eq. (\ref{eq:mass}) and assuming $\gamma = 5/3$,
\begin{eqnarray}
	\vout &\sim& \sqrt{\frac{2(\gamma-1)}{6\gamma-1}} \sqrt{ \frac{\Bin^2}{\mu_0\rhoin} + \frac{\vin^2}{2} } \nonumber \\
	&\sim& 0.38 \sqrt{v_A^2 + \frac{\vin^2}{2}}. \label{eq:vout}
\end{eqnarray}
Here, $v_A$ is the Alfven velocity defined by 
$\Bin$ and $\rhoin$ in the upstream region.
	As observed in the SE image at $t=7$ ns [Fig. \ref{fig:se}(c)], the inflow plasmas
stagnate near the mid-plane, suggesting
$\vin \sim 0$ and $\ekin \sim 0$.
Therefore, Eq. (\ref{eq:vout}) becomes
\begin{eqnarray}
	\vout \sim 0.38 v_A. \label{eq:vout2}
\end{eqnarray}
The acceleration of ion flows observed at $t=7$ ns shown in 
Figs. \ref{fig:TS_para_vi} and \ref{fig:TS_para_param}(c) 
can be interpreted as the outflows
accelerated by magnetic reconnection.
Using $\vout = 30$ km/s ($v_A = 78$ km/s),
$\NI\sim1.5\times10^{18}$ cm$^{-3}$ (half of the measured ion density),
$\te\sim70$ eV, $\ti\sim100$ eV, and $Z\sim5.6$
for lower-temperature component (Fig. \ref{fig:TS_para_param}), and
Eqs. (\ref{eq:Binj}), (\ref{eq:jB}), and (\ref{eq:vout2}),
$\Bin \sim 15$ T, 
$\delta_e \sim 11$ $\um$, and
$j_z\sim 2.2\times10^{12}$ Am$^{-2}$.
The current 
is roughly consistent with
other previous measurements with proton radiography.
For example, in the experiment at National Ignition Facility
(NIF)\cite{Fox2020-xb},
path-integrated peak current density was estimated as
$\int j_z dl\sim1.6\times10^{8}$ Am$^{-1}$
or $1.6\times10^{11}$ Am$^{-2}$ with a measured magnetic field of
$\Bin\sim5$ T assuming the integrated length of 1 mm,
averaged in the thickness of 58 $\um$, which is comparable to
the present result, $2.2\times10^{12}$ Am$^{-2}$ in the electron
current sheet of 11 $\um$ or 
$\sim2.4\times10^{11}$ Am$^{-2}$ averaged 
in the measurement region of $\sim$100 $\um$.

The magnetic field diffusion rate in the current sheet 
or reconnection rate 
is the rate of the field flux change in the upstream region:
\begin{eqnarray}
	R = \frac{1}{v_A \Bin}\left|\frac{d\Phi_{\rm in}}{dt}\right|,
\end{eqnarray}
where the rate is normalized by $v_A\Bin$, 
$\Phi_{\rm in}$ is 
the magnetic field flux in the upstream region 
$\Phi_{\rm in}=\int B_x dy$ in the $xy$-plane assuming 
$\partial B_x/\partial z = 0$,
and is
roughly $\Phi_{\rm in}\sim\Bin\lambda_B/2$ assuming 
$B_x(y) \sim -B_{\rm in} (y/\lambda_B)$
in the plasma sheet,
and
$\Bin\sim15$ T is estimated at $t=7$ ns.
As the asymmetric spectra, suggesting the ion flow acceleration, 
appears at $t=$ 5--7 ns 
(see Figs. \ref{fig:TS_para} and \ref{fig:TS_para_param})
and the current disappears at $t=9$ ns (see Fig. \ref{fig:current}).
Although $\lambda_B$ and $\Bin$ change during the magnetic reconnection and
we have no magnetic field measurement here,
we assume $\lambda_B \sim \rci \sim 60$ $\um$ and
total flux, $\Phi_{\rm in}$ disappears
in $\Delta t \sim 4$ ns (5--9 ns).
Therefore, the reconnection rate becomes
$R\sim0.096$,
which is comparable to
the universal reconnection rate of 0.1
for Hall magnetic reconnection where the rate is determined
only by local plasma parameters\cite{Huba2004-tl,Cassak2017-dw,Yamada2010-jd}.
However, the rate would be precisely determined by measuring 
a spatial distribution of $\Bin$ and $\lambda_B$ by
proton radiography technique\cite{Fox2020-xb,Li2007-hy,Fox2012-wm}
in addition to multi-directional LTS measurements
in future experiments.
Though $\Bin$ is estimated only at $t=7$ ns in the present experiment,
the time-evolution of the rate would also be determined 
by measuring the time-evolution of LTS spectra in both directions
with better time-resolution or with a streak camera instead of ICCD camera.

For the estimation of plasma parameters in the $\vector{k_1}$ direction, 
$\sim$87\% of the ions are the high-temperature component 
while only $\sim$13\% are
the low-temperature component, i.e. $\rhohot/\rhocold=0.87/0.13=6.7$,
where $\rhohot$ and $\rhocold$ 
are the densities of the high- and low-temperature components, respectively.
Here, the high-temperature component is considered as the outflow
from magnetic reconnection, 
\begin{eqnarray}
	\rhohot \sim \rhoout, \label{eq:rhoout}
\end{eqnarray}
while the low-temperature component is the plasma
coming from the two laser-spots directly.
However, as shown in Fig. \ref{fig:TS_para_param}(a),
the density of the low-temperature component is smaller than 
that of the single flow, i.e., $\rhocold < \rhosingle$.
In addition, $\rhoin$ should be larger than $\rhosingle$
because of the deceleration of the inflows as observed in SE images,
resulting in 
\begin{eqnarray}
	\rhoin > \rhosingle > \rhocold = r\rhoin \hspace{10pt} (0<r<1). \label{eq:rhoin}
\end{eqnarray}
From the conservations of magnetic field flux and mass,
\begin{eqnarray}
	\frac{\rhoout}{\rhoin} = \frac{\vin}{\vout}\frac{L}{\delta} \sim R\frac{L}{\delta}, \label{eq:rhoratio}
\end{eqnarray}
where
$R=\vin/\vout = \Bout/\Bin$ is the reconnection rate which is 
already estimated as $R\sim 0.1$.
The length $L$ is, generally, larger than $\delta$, 
and we observed bidirectional flows separated by $\sim$400 $\um$ 
along $p$-axis [see Fig. \ref{fig:TS_para_param}(b)] 
indicating $\sim$300 $\um$ along $x$-axis, and roughly $L\sim300$ $\um$,
while $\delta \sim 60$ $\um$.
Therefore, using Eqs. (\ref{eq:rhoout})--(\ref{eq:rhoratio}), we get
\begin{eqnarray}
	\frac{\rhohot}{\rhocold} = \frac{\rhoout}{r\rhoin} 
	\sim \frac{RL}{r\delta} \sim 0.5/r,
\end{eqnarray}
and the observed ratio $\rhohot/\rhocold = 6.7$ is explained with small $\rhocold$ 
relative to $\rhoin$
or small $r \sim 0.075$.

Here, the ion and electron Larmor radii are estimated as
$\rci\sim60$ $\um$ and $\rce\sim1.9$ $\um$, respectively, at $t=7$ ns,
(using $\ti=100$ eV, $\te=70$ eV, $\zi=5.6$, and $B=15$ T)
and both are smaller than typical system size $l_{\rm typ}\sim1$ mm.
In addition, the spatial resolution of the LTS is roughly 
$l_{\rm LTS}\sim100$ $\um$
[see LTS volume in Fig. \ref{fig:setup}(d)],
and
$\rce \ll \rci \lesssim l_{\rm LTS} \ll l_{\rm typ}$, 
meaning that both electrons and
ions are magnetized in the ablation plasma,
and electrons are still magnetized in the LTS measurement volume.
As previously reported\cite{Li2007-hy}, the laser-produced plasma has
large $\te$ ($\sim1$ keV) and $\NE$ ($\sim10^{20\mathchar`-\mathchar`-22}$
cm$^{-3}$) 
early in time around the laser-spot, and strong magnetic field
of $B\sim100$ T is formed via Biermann battery effect,
resulting in
large plasma beta, $\beta_e = 2\mu_0\NE\te/B^2\sim$ 4--400.
It becomes small, for example, at $t\sim7$ ns, $\beta_e\sim1.3$.
The Lundquist number at $t\sim7$ ns is estimated as
$S=Lv_A/D_M\sim 270$, where $L$ is the system size of $\sim$1 mm,
$v_A = \Bin/\sqrt{\mu_0\rhoin}\sim 78$ km/s
in the upstream region,
$D_M = \nu_{ei}(c/\omega_{pe})^2 \sim 0.29$ m$^2$s$^{-1}$ 
is the magnetic diffusivity,
and $\nu_{ei} = 1.2\times10^{11}$ s$^{-1}$
is electron--ion collision frequency\cite{Chen2016-oi}
with $\NE\sim1.7\times10^{19}$ cm$^{-3}$ and $\te\sim70$ eV.
Previous numerical simulations\cite{Biskamp1986-tp,Bhattacharjee2009-rq,Samtaney2009-ak} 
have suggested that the current sheet becomes
stochastic in high Lundquist number such as $S>10^4$
due to tearing instability,
but is in quasi-steady state in relatively small $S$ $(<10^4)$
which is in our experimental condition.

The spectra parallel to $\Bin$ (Fig. \ref{fig:TS_para}) show 
sharp peaks suggesting ion-acoustic resonance.
On the other hand, the spectra perpendicular to $\Bin$ 
(Fig. \ref{fig:TS_perp_spec}) show
weak resonance suggesting strong damping of ion-acoustic waves
with $\ti > \te$.
These difference in spectra are explained with non-equal temperature in two directions:
$\tiperp \ne \tipara$ and/or $\teperp \ne \tepara$, 
where $\parallel$ and $\perp$ represent the directions
relative to $\Bin$, respectively.
This anisotropy in velocity distribution 
may come from Speiser orbits and meandering motions around the diffusion region
and has been observed by
Magnetospheric Multiscale Mission (MMS) observation\cite{Wang2016-sd}
and numerical simulations\cite{Zenitani2013-bk}, and can be
further investigated in future experiment in the method presented here.

In the present experiment, 
the Biermann battery fields are advected with expanding plasmas
and the anti-parallel field structure is formed in the $x$ direction
near the plane $y=0$ and the reconnection can occur
anywhere in the $z$-axis.
As the SE imaging shows,
two plasmas from the top and bottom interact with each other
at $z\sim0$,
while the anti-parallel field structure still exists at $z\gtrsim0$
at $t=7$ ns.
This indicates that the magnetic pressure decreases at $z\sim0$
due to magnetic field diffusion in the current sheet.
The electron current is measured from $t=5$ ns to 9 ns
in the $z$ direction, accompanied by the bipolar ion flows
accelerated in the $\pm x$ directions at $t=5$ and 7 ns.
These measurements suggest that the anti-parallel Biermann fields
reconnect in the electron current sheet at $(x,y,z) \sim (0,0,0)$,
accelerating the plasma as outflows during $t\sim$ 5--9 ns,
and it ends at $t\gtrsim9$ ns.

\section{Summary}
\label{sec:sum}

We have measured the appearance and disappearance of 
an electron current sheet as well as bidirectional ion flows,
for the first time,
in the magnetic reconnection between laser-produced magnetized plasmas.
We have investigated magnetic reconnection and magnetic diffusion region formed
in a self-generated anti-parallel magnetic field by using optical
diagnostics: two-directional laser Thomson scattering and
self-emission imaging.
Thomson scattering spectra perpendicular to the magnetic field
show different Landau damping effects on ion-acoustic waves
in the $\pm z$ directions, indicating the current sheet formation.
The spectra parallel to the magnetic field show different widths
in two peaks, which are interpreted as non-Maxwellian ion velocity distribution
and two different components: cold and slow ions and hot and faster ions.
This acceleration along the magnetic field is explained as the outflow
from the magnetic reconnection.
Assuming that the additional velocity is comparable to the Alfven velocity
defined with the upstream plasma parameters,
the magnetic field in the upstream region is $\Bin\sim15$ T.
The current density in the electron current sheet 
$j_z \sim2.2\times10^{12}$ Am$^{-2}$ is nearly consistent with that obtained
from the spectra obtained in the perpendicular direction: 
$j_z \sim$ (0.4--1.9)$\times10^{12}$ Am$^{-2}$, and is also
comparable to the estimation from proton radiography of
similar laser experiments.
Combining two directional data, the current is formed from
$t\sim5$ to 9 ns accompanied by bidirectional plasma flows 
observed at $t\sim7$ ns due to magnetic reconnection.
The SE imaging shows the stagnation of two plasmas
showing two separated dense regions. These structures interact
and merge at $t>7$ ns, which can be interpreted as the magnetic pressure
decrease due to magnetic reconnection.
While the electron current is detected here,
both electron and ion velocity distributions are needed
to directly measure the current sheet.
These can be measured with non-collective LTS for
smaller density or shorter wavelength for the probe laser,
or the electron velocity distribution at the phase velocity of 
electron plasma waves can be obtained from the electron-feature
of collective LTS in future experiments.
Also, multiple-direction LTS can reveal asymmetric ion velocity distributions
in $x$, $y$, and $z$ directions resulting from 
Speiser orbits and meandering motions in the outflow region,
which has been analyzed and discussed
by using particle-in-cell simulations\cite{Zenitani2013-bk}
and by MMS observation\cite{Wang2016-sd}.

The magnetic reconnection rate is estimated as $R\sim 0.1$
assuming spatial distribution of $\Bin(y)$ averaged in 5--9 ns.
This reconnection rate can be estimated precisely 
by using LTS presented here
and by using magnetic field measurement, for example, proton radiography
simultaneously
in future experiments.

\begin{acknowledgments}
	The authors would like to acknowledge the dedicated technical
	support of the staff at the Gekko-XII facility for the laser operation,
	target fabrication, and plasma diagnostics. 
	We would also like to thank N. Ozaki for the target alignment 
	in the experiment, and M. Hoshino, S. Zenitani, Y. Ohira, and N. Yamamoto
	for helpful comments and valuable discussions. 
	This research was partially
	supported by JSPS KAKENHI Grant Nos. 22H01251, 20H01881, 18H01232, and 17H06202,
	and by the joint research
	project of Institute of Laser Engineering, Osaka University
\end{acknowledgments}

%






%

\end{document}